\def\papertitle{Differentiable grey-box modelling of phaser effects using frame-based spectral processing}
\def\paperauthorA{Alistair Carson}
\def\paperauthorB{Cassia Valentini-Botinhao}
\def\paperauthorC{Simon King}
\def\paperauthorD{Stefan Bilbao}

\documentclass[twoside,a4paper]{article}
\usepackage{etoolbox}
\usepackage{dafx_23}
\usepackage{amsmath,amssymb,amsfonts,amsthm}
\usepackage{euscript}
\usepackage[T1]{fontenc}
\usepackage[utf8]{inputenc}
\usepackage{ifpdf}
\usepackage[english]{babel}
\usepackage{caption}
\usepackage{subfig} 
\usepackage{color}
\usepackage{enumitem}
\usepackage{ifsym}

\usepackage[dvipsnames]{xcolor}

\def\SB[#1]{\textcolor{red}{#1}}
\def\AC[#1]{\textcolor{blue}{#1}}

\input glyphtounicode
\ninept

\newcounter{numauth}
\newcounter{listcnt}
\newcommand\authcnt[1]{\ifdefined#1 \stepcounter{numauth} \fi}

\newcommand\addauth[1]{
\ifdefined#1 
\stepcounter{listcnt}
\ifnum \value{listcnt}<\value{numauth}
\appto\authorslist{, #1}
\else
\appto\authorslist{~and~#1}
\fi
\fi}
\authcnt{\paperauthorB}
\authcnt{\paperauthorC}
\authcnt{\paperauthorD}
\authcnt{\paperauthorE}
\authcnt{\paperauthorF}
\authcnt{\paperauthorG}
\authcnt{\paperauthorH}
\authcnt{\paperauthorI}
\authcnt{\paperauthorJ}
\def\authorslist{\paperauthorA}
\addauth{\paperauthorB}
\addauth{\paperauthorC}
\addauth{\paperauthorD}
\addauth{\paperauthorE}
\addauth{\paperauthorF}
\addauth{\paperauthorG}
\addauth{\paperauthorH}
\addauth{\paperauthorI}
\addauth{\paperauthorJ}

\usepackage{times}

\newif\ifpdf
\ifx\pdfoutput\relax
\else
   \ifcase\pdfoutput
      \pdffalse
   \else
      \pdftrue
\fi

\ifpdf 
  \usepackage[pdftex,
    pdftitle={\papertitle},
    pdfauthor={\authorslist},
    pdfsubject={Proceedings of the 26th International Conference on Digital Audio Effects (DAFx23)},
    colorlinks=false, 
    bookmarksnumbered, 
    pdfstartview=XYZ 
  ]{hyperref}
  \usepackage[pdftex]{graphicx}
\else 
  \usepackage[dvips]{epsfig,graphicx}
  \usepackage[dvips,
    pdftitle={\papertitle},
    pdfauthor={\authorslist},
    pdfsubject={Proceedings of the 26th International Conference on Digital Audio Effects (DAFx23)},
    colorlinks=false, 
    bookmarksnumbered, 
    pdfstartview=XYZ 
  ]{hyperref}
\fi
\usepackage[hypcap=true]{caption}
\usepackage{siunitx}
\title{\papertitle}

\fouraffiliations{
\paperauthorA \,\sthanks{A Carson is funded by the Scottish Graduate School of Arts and Humanities (SGSAH).}}
{\href{http://www.acoustics.ed.ac.uk}{Acoustics and Audio Group} \\ University of Edinburgh \\ Edinburgh, UK\\
{\tt \href{mailto:alistair.carson@ed.ac.uk}{alistair.carson@ed.ac.uk}}
}
{\paperauthorB \,}
{\href{https://www.cstr.ed.ac.uk}{Centre for Speech Technology Research} \\ University of Edinburgh\\ Edinburgh, UK\\ {\tt \href{mailto:cvbotinh@inf.ed.ac.uk}{cvbotinh@inf.ed.ac.uk}}
}
{\paperauthorC \,}
{\href{https://www.cstr.ed.ac.uk}{Centre for Speech Technology Research} \\ University of Edinburgh\\ Edinburgh, UK\\ {\tt \href{mailto:simon.king@ed.ac.uk}{simon.king@ed.ac.uk}}
}
{\paperauthorD \,}
{\href{http://www.acoustics.ed.ac.uk}{Acoustics and Audio Group} \\ University of Edinburgh\\ Edinburgh, UK\\ {\tt \href{mailto:sbilbao@ed.ac.uk}{sbilbao@ed.ac.uk}}
}

\begin{document}
\ifpdf 
  \DeclareGraphicsExtensions{.png,.jpg,.pdf}
\else  
  \DeclareGraphicsExtensions{.eps}
\fi


\maketitle

\begin{abstract}
Machine learning approaches to modelling analog audio effects have seen intensive investigation in recent years, particularly in the context of non-linear time-invariant effects such as guitar amplifiers. For modulation effects such as phasers, however, new challenges emerge due to the presence of the low-frequency oscillator which controls the slowly time-varying nature of the effect. Existing approaches have either required foreknowledge of this control signal, or have been non-causal in implementation. This work presents a differentiable digital signal processing approach to modelling phaser effects in which the underlying control signal and time-varying spectral response of the effect are jointly learned. The proposed model processes audio in short frames to implement a time-varying filter in the frequency domain, with a transfer function based on typical analog phaser circuit topology. We show that the model can be trained to emulate an analog reference device, while retaining interpretable and adjustable parameters. The frame duration is an important hyper-parameter of the proposed model, so an investigation was carried out into its effect on model accuracy. The optimal frame length depends on both the rate and transient decay-time of the target effect, but the frame length can be altered at inference time without a significant change in accuracy.


\end{abstract}

\section{Introduction}
\label{sec:intro}
A broad class of audio effects found in almost all genres of popular music is that of time-varying modulation effects, and includes phasing, flanging, chorus, and tremolo. Model-based digital implementations of these effects are straightforward \cite{DutilleuxDAFx}, but many musicians prefer the timbre and character of the original analog or electro-mechanical devices used to create these effects, and these may be considerably more difficult to model. Circuit-based simulations \cite{holters2011, Eichas2014} can produce physically accurate results but are highly optimized to a specific device, and require complete knowledge of the circuit and its component values. 

In general, modelling of audio effects using machine learning has become an active area of research in recent years, with a particular focus on modelling non-linear time-invariant effects such as guitar amplifiers and distortion pedals. Approaches in modelling these systems include fully black-box methods using recurrent or convolutional neural networks (RNNs / CNNs) \cite{Damskaag2018, Wright2020}, as well as grey-box models which use some prior knowledge of the reference system, such as differentiable state-space models \cite{Parker2019, Esqueda2021} and differentiable DSP-based models \cite{Kuznetsov2020, Nercessian2021}. Modelling of effects with time-varying input-dependent behaviour such as dynamic range compression has also been explored through the use of CNNs with long receptive fields \cite{steinmetz2022} and temporal feature-wise linear modulation \cite{Comunita2022}; as well as a differentiable DSP model proposed in \cite{Wright2022}. Modelling of time-varying modulation effects presents a unique challenge due to the modulation of system behaviour by a low frequency oscillator (LFO). The deep-learning approach proposed in  \cite{Ramirez2020_BB} can emulate a wide range of effects, however the use of bi-directional long-term short-memory networks (LSTMs) makes the model non-causal and therefore not suited for real-time use. Wright et al. \cite{Wright2021} proposed a real-time model for phasing and flanging effects using RNNs, but required manual measurement and estimation of the LFO prior to training as this was an input to the model. Earlier work by Kiiski et al. \cite{Kiiski2016} proposed a grey-box model of phasing, but in which no machine learning was used.

This work presents a differentiable DSP model of a phaser effect that can be trained through gradient descent to jointly learn the underlying LFO signal and the time-varying spectral response of an analog reference pedal. The model utilises frequency domain approximations of IIR filters to accelerate training times, as has been employed in \cite{Nercessian2021, Wright2022}. Through experiment, we investigate the conflicting demands of time and frequency resolution when using this method in the context of time-varying effects.

The paper is structured as follows: Section \ref{sec:phasing} provides background on analog and digital phasing effects; Section \ref{sec:model} outlines the proposed model; Section \ref{sec:target_systems} describes the target systems and data; Section \ref{sec:experiments} describes the experimental procedure and results; and Section \ref{sec:conclusion} concludes the paper with an outlook on areas of future work. 
Source code and audio examples are provided at the accompanying web-page \cite{webpage}.

\section{Analog and Digital Phase Shifters}\label{sec:phasing}
The phaser is a time-varying filter effect in which the phase of an input signal is subject to a periodic modulation and combined together with the `dry' (uneffected) signal to create audible notches in the frequency spectrum \cite{bartlett1970a}. The movement of these notches with time gives the characteristic perceived sweeping effect. The phasing effect is often confused with flanging, and indeed both effects are caused by modulating notches in the spectra. The key difference is that a flanger generates an infinite series of harmonically spaced notches, whereas the phaser response has finite non-uniformly spaced notches \cite{PASPWEB2010}. Furthermore, implementations of phasers and flangers differ both historically and in present day commercial products. Originally, the flanging effect was created by playing two tape machines in near unison with a small variable time delay between the two reels \cite{bode1984history}. Nowadays, flangers are typically implemented using time-varying digital comb filters, or `bucket-brigade' delay lines in analog pedals. In contrast, phasers create phase shifts through cascaded all-pass filters, in which break frequencies are modulated by a low-frequency oscillator (LFO). 

Digital phaser effects are usually implemented as linear time-varying signal-based models, in which the all-pass filters are discrete-time approximations of idealised all-pass filters \cite{DutilleuxDAFx, PASPWEB2010, DAFX_VAchapter}. The motivation of this work is to explore whether such a model can be embedded in a machine learning framework to emulate the response of an analog phaser pedal. The continuous and discrete time signal processing concepts of phasing, which underpin the proposed model, are outlined in the remainder of this section.


\subsection{Continuous-time phasing}
\begin{figure}[ht]
\centerline{\includegraphics[scale=0.7, clip, trim=3mm 0mm 3mm 0mm]{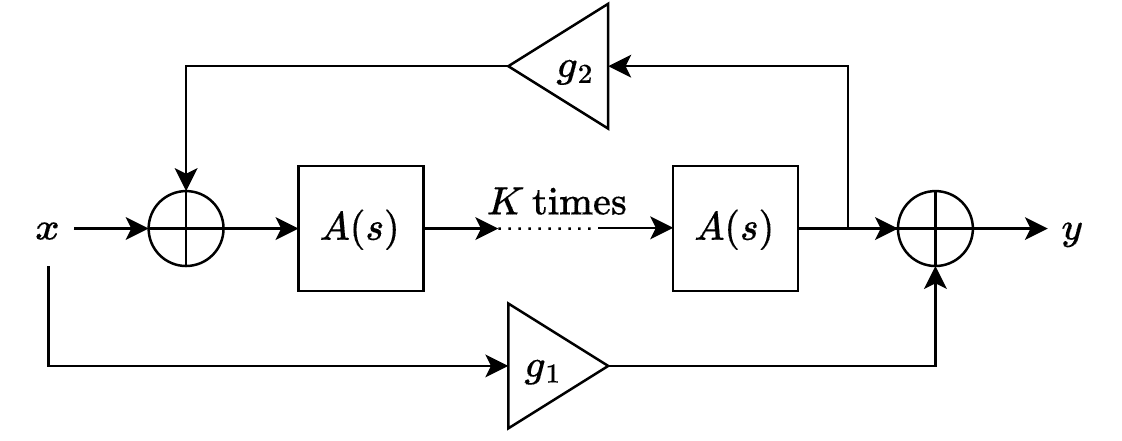}}
\caption{\label{s_domain_block}{\it A typical phaser structure in continuous time.}}
\end{figure}

We will examine phasers of the topology shown in Figure \ref{s_domain_block} composed of a cascade of $K$ identical first-order all-pass filters. (In many practical implementations of the phaser, $K=4$ \cite{PASPWEB2010}, but other choices of $K$ have been employed \cite{Kiiski2016}.) The continuous-time transfer function of each such first-order section is \cite{PASPWEB2010}:
\begin{equation}\label{eq:H_ap(s)}
    A(s) = \frac{s - \omega_b}{s + \omega_b}
\end{equation}
where $s = \sigma + j \omega$ is the complex frequency and $\omega_b\geq 0$ is known as the \textit{break-frequency} of the all-pass filter. 
Because $A$ is all-pass, for real frequencies $s=j\omega$ it may be written as $A = e^{j\Theta}$, where:
\begin{equation}\label{ap-phase-respose-s}
    \Theta(\omega) = \pi - 2\arctan \left( \omega/\omega_{b}\right)\,.
\end{equation}
From \eqref{ap-phase-respose-s} it can be observed that $\omega_b$ is the frequency at which $\Theta(\omega_b) = \pi / 2$. Furthermore, the all-pass filter inverts DC, and the phase response tends to zero as $\omega \to \infty$. 

\begin{figure}[t]
\centerline{\includegraphics[scale=0.62,clip,trim=40mm 98mm 40mm 98mm]{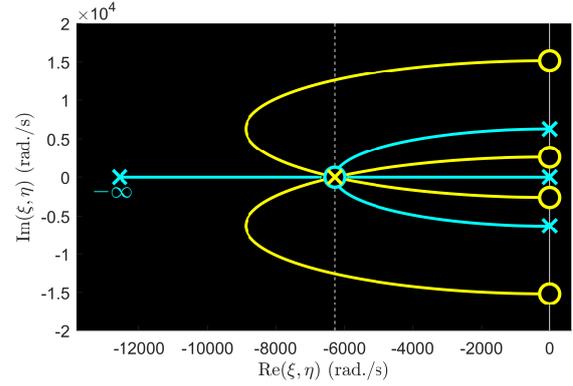}}
\caption{\label{rlfig}{\it Root locus plot, for $g_{1} = 1$, and for for $g_{2}$ between 0 (open loop, indicated by circles) and 1 (indicated by crosses). Pole locations are indicated by blue lines, and zero locations by yellow lines. The real frequency $-\omega_{b}$ (where here, $\omega_{b} = 2\pi\cdot 1000$ rad/s) is indicated by a dashed line.  }}
\end{figure}

\begin{figure}[ht]
\centerline{\includegraphics[scale=0.67,clip,trim=38mm 106mm 40mm 107mm]{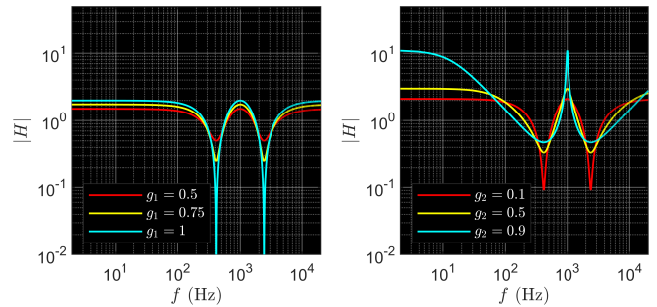}}
\caption{\label{phaser_tf_g2}{\it Magnitude response for $H$, under open-loop conditions at left, for different values of $g_{1}$, as indicated, and under closed-loop conditions at right, for $g_{1} = 1$, and different values of $g_{2}$, as indicated. As before, $\omega_{b} = 2\pi\cdot 1000$ rad/s.}}
\end{figure}

The cascade arrangement in Figure \ref{s_domain_block} includes a through path  with gain $g_1$, and a feedback path with gain $g_2$, with $0\leq g_{2}<1$. The following transfer function results:
\begin{equation}\label{eq:phaser_tf_s}
    H(s) = g_1 + \frac{A^K}{1 - g_2 A^K}\,.
\end{equation}
The poles $s=\xi_{k}$, and zeros $s=\eta_{k}$, $k=0\hdots,K-1$, are:
\begin{subequations}
\begin{eqnarray}
\hspace{-0.4in}\xi_k \!\!\!\!&=&\!\!\!\! \omega_b \frac{1 \!+\! \lambda_k}{1 \!-\! \lambda_k}\quad {\rm where}\; \lambda_{k} = \frac{e^{j2\pi k/K}}{\sqrt[K]{g_{2}}}\\
\eta_k \!\!\!\!&=&\!\!\!\! \omega_b \frac{1 \!+\! \beta_k}{1 \!- \!\beta_k}\quad {\rm where}\; \beta_{k} = \sqrt[K]{\frac{g_1}{1 \!-\! g_1g_2}} e ^ {j \pi (2 k + 1)/ K} \,.
\end{eqnarray}
\end{subequations}
Figure \ref{rlfig} shows a root locus plot of the pole and zero trajectories for $0\leq g_{2}\leq 1$, and when $g_{1} = 1$. Under open loop ($g_{2} = 0$) conditions, two notches are located on the non-negative imaginary axis.  As $g_1 \to 0$ the zeros move away from the imaginary axis  and as such, parameter $g_1$ controls the perceived depth of the phaser effect. Under fully closed loop conditions, two poles are located on the non-negative imaginary axis (one at DC).Variations in the transfer function magnitude $|H|$ with both $g_{1}$ and $g_{2}$ are shown in Figure \ref{phaser_tf_g2}. 

\subsection{Discrete-time phasing}

\begin{figure}[ht]
\centerline{\includegraphics[scale=0.7]{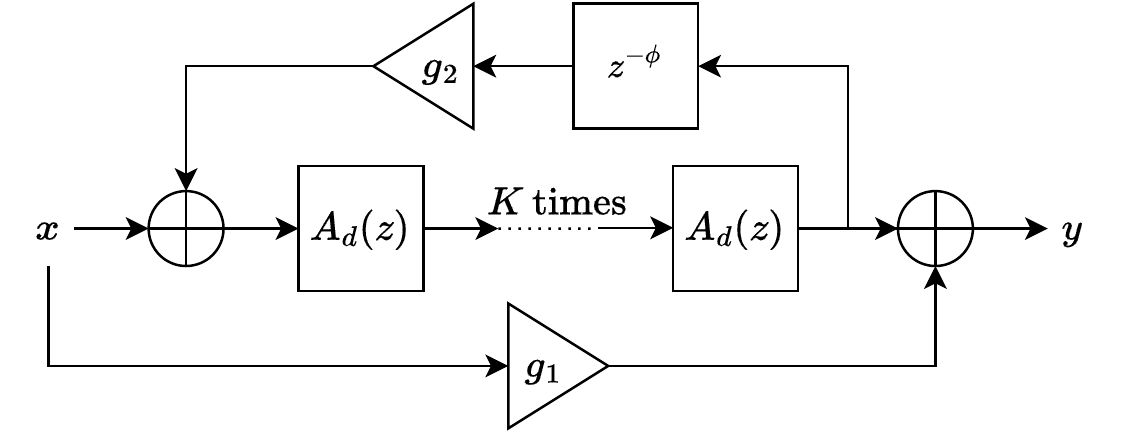}}
\caption{\label{fig:td_model}{\it A typical DSP phaser structure. }}
\end{figure}

\begin{figure}[ht]
\centerline{\includegraphics[scale=0.61,clip,trim=38mm 103mm 40mm 103mm]{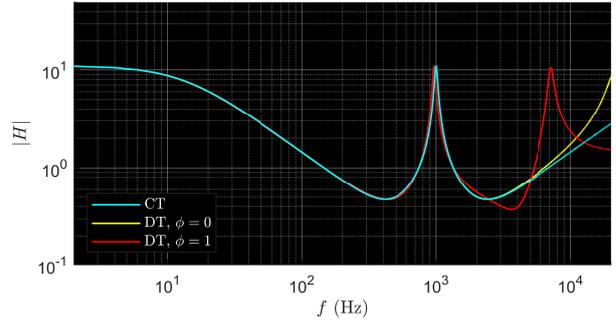}}
\caption{\label{fig:tf_compare}{\it Comparison of magnitude responses of continuous-time (CT) and discrete-time (DT) phaser models, with constant parameters $\omega_b = 2\pi\cdot 1000$ rad/s, $g_1 = 1.0$, $g_2 = 0.9$, and $T = 1/44100$ s. Distinct cases of the discrete-time model without ($\phi = 0$) and with ($\phi = 1$) a fictitious delay are illustrated. }}
\end{figure}

The continuous-time transfer function \eqref{eq:H_ap(s)} can be approximated in discrete-time via the bilinear transform to give the discrete-time all-pass section, $A_{d}(z)$, defined as:
\begin{equation}\label{eq:A(z)}
    A_{d}(z) = \frac{p - z^{-1}}{1 - p z^{-1}}\qquad {\rm where}\qquad p = \frac{1 - \tan(\omega_b T /2)}{1 + \tan(\omega_b T /2)}\,.
\end{equation}
Here, $T = 1/F_{s}$ is the sampling period, for sample rate $F_{s}$ in Hz. The corresponding discrete-time phaser is shown in Figure \ref{fig:td_model}. The discrete-time transfer function $H_{d}(z)$ of $K$ cascaded all-pass sections, including the residual connection and feedback loop is:
\begin{equation}\label{eq:tf_z-domain}
    H_{d}(z) = g_1 + \frac{A_{d}^{K}}{1 - g_2 z ^ {-\phi} A_{d}^{K}}\,.
\end{equation}
Note that a digital delay $z^{-\phi}$, $\phi \in \mathcal{Z}^{+}$ has been included in the feedback loop. Without this, a delay-free loop would be present in the resulting difference equation. Efficient modelling of delay-free loops found in phaser pedals has proven challenging: the state-space model of a phaser pedal presented in \cite{Eichas2014} required a Newton-Raphson solver at run-time; and Kiiski et al. reported that the fictitious delay line in their DSP model ($\phi=1$) resulted in perceptual differences in the feedback effect when compared with the analog reference device \cite{Kiiski2016}. The effect of the delay line on the magnitude response of the system can be seen in Figure \ref{fig:tf_compare}. It is clear that the fictitious delay line in the feedback loop drastically changes the magnitude response for high-frequencies, and should ideally be avoided in virtual analog models of phasers.

\subsection{Time-varying behaviour}\label{subsec:lfos}
In analog phase shifters, the break frequency of the all-pass sections, $\omega_b$, is periodically modulated by a sub-audio rate LFO, typically with frequency in the range \SI{0.05}{\hertz} to \SI{5}{\hertz}. The mapping between LFO voltage and break frequency is often nonlinear and asymmetrical: for example, in the MXR Phase 90 the LFO varies the voltage across JFETs which in turn alters the break frequency of the all-pass \cite{Eichas2014}. Certain devices, such as the Uni-Vibe and its replicas, are optical: the LFO varies the voltage across a light source (an incandescent bulb in original units \cite{Darabundit2019}) surrounded by light dependent resistors (LDRs). The resulting fluctuation in current through the LDRs controls the all-pass break frequencies. Accurate prediction of the LFO characteristics (including fundamental frequency and waveshape) are therefore critical to the quality of discrete-time emulations of such effects, including phasers. 

\section{Modelling Method}\label{sec:model}
\begin{figure}[h]
\center
\includegraphics[scale=0.7]{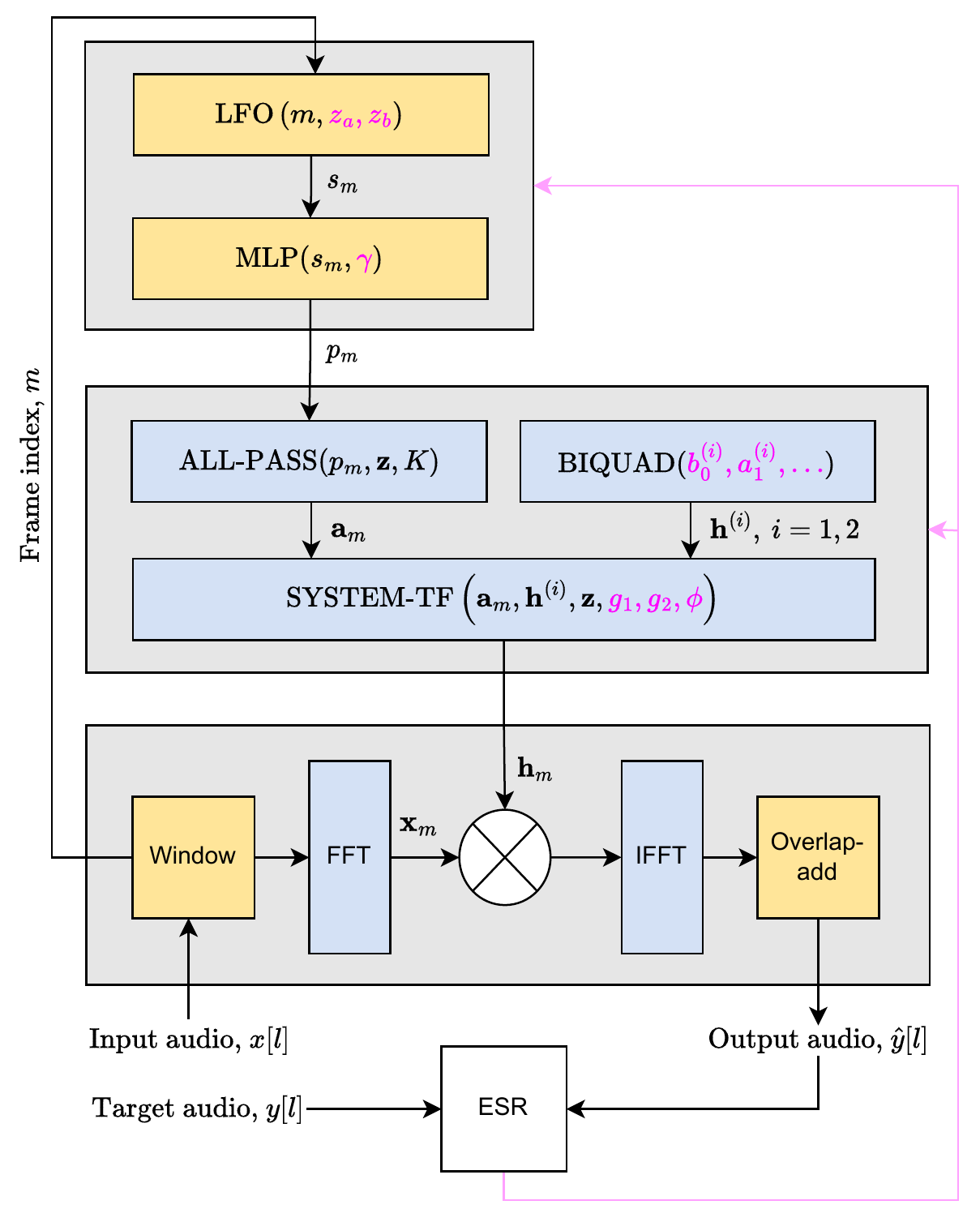}
\caption{\label{fig:proposed_model}{\it Structure of the proposed model. Black arrows indicate signal flow, with magenta indicating the flow of gradients to the learnable parameters. The blue and yellow boxes indicate operations in the frequency domain and time domain respectively. The subscript $m$ denotes parameters varying at the frame rate, with all other parameters held constant for the duration of a training epoch. Parameters in magenta are updated once per epoch by the optimizer.}}
\end{figure}

This section outlines the proposed model of a differentiable phase shifting algorithm which, given input-output audio recordings of a reference device, can be trained to emulate the time-varying behaviour. Consider an arbitrary phase-shifting device that has been sampled in discrete time as:
\begin{equation}\label{eq:target_sytem}
    y[n] = f(x[n], \theta[n])
\end{equation}
where $x$ is the input signal, $y$ is the output signal, $\theta$ are the time-varying parameters of the system, $n$ is the sample index and $f$ is a linear function of $x$. We seek to develop a model of the form:
\begin{equation}\label{eq:high_level_model}
    \hat{y}[n] = g(x[n], \hat{\theta}[n])
\end{equation}
whose output $\hat{y}[n]$ is perceptually indistinguishable from $y[n]$. The function $g$ is assumed differentiable so that the model can be trained using gradient descent to find model parameters $\hat{\theta}$ that minimise an objective loss function $\mathcal{L}(y, \hat{y})$. The proposed model architecture is shown in Figure \ref{fig:proposed_model}.

\subsection{Frame-based processing}
Utilising frame-based spectral processing \cite{stockham1966, BonadaDAFx}, the proposed model assumes that the target phaser can be treated as a linear time-invariant (LTI) system over the duration of a short frame of length $W$ seconds. Suppose that an input audio signal $x[l]$ of length $L$ samples is segmented into $N_{f} = \lceil L/H\rceil$ frames of length $N = \lfloor W F_s \rfloor$ with hop-size $H$ samples. (The resulting frame rate is $F_f = F_s / H$ in Hz.) The $m$th frame, $m=0,\hdots,N_{f}-1$, is defined as the $N\times 1$ column vector ${\bf x}_{m} = [x[mH],\hdots,x[mH+N-1]]^{T}$. The short-time Fourier transform vector ${\bf X}_{m}$, $m=0\hdots,N_{f}-1$ is derived from ${\bf x}_{m}$ through windowing and Fourier transformation as follows:
\begin{equation}
    {\bf X}_{m} = {\bf U}{\bf Q}{\bf x}_{m} \,.
\end{equation}
Here, ${\bf Q} = [{\bf W} \,\, {\bf Z}]^{T}$ is an $N'\times N$ windowing matrix, where where $N'$ is the DFT length. It includes the diagonal $N\times N$ matrix ${\bf W}$, containing samples of the Hann window on the diagonal, and an $N\times (N'-N)$ all-zero matrix ${\bf Z}$ implementing zero-padding. ${\bf U}$ is an $N'\times N'$ DFT matrix. 
At each frame, spectral processing is applied via element-wise complex multiplication in the frequency domain, followed by an inverse Fourier transform, truncation and windowing to yield the $N\times 1$ output vectors ${\bf y}_{m}$, $m=0,\hdots,N_{f}-1$:
\begin{equation}\label{eq:spectral_process}
    {\bf y}_{m} = \frac{1}{N'}{\bf Q}^{T}{\bf U}^{*}{\bf H}_{m}{\bf X}_{m} \,.
\end{equation}
Here, $\mathbf{H}_m$ is an $N'\times N'$ diagonal matrix containing values of a transfer function (incorporating Hermitian symmetry) at frame $m$ on its diagonal. ${\bf U}^{*}$ is the conjugate transpose of the DFT matrix ${\bf U}$. Finally, the output time series $\hat{y}[l]$ is obtained from the frames ${\bf y}_{m}$ through an overlap-add procedure, with hop size $H$. It can be shown that for $\mathbf{H}_m = {\bf I}$, exact reconstruction of the input signal can be obtained if $N/H$ is an integer greater than two---this property is known as constant-overlap-add and is enforced in the proposed model. Considering the transfer function in \eqref{eq:spectral_process} to represent a digital filter with  $M \in \mathbb{Z}$ non-zero taps, then the DFT length must be $N' \geq N + M - 1$ to avoid temporal aliasing in the output frames. For IIR filtering (where $M \to \infty$), some degree of temporal aliasing is inevitable but can be practically suppressed by choosing $M$ as the 60dB decay time (in samples) of the filter. In this work the decay time of the target system was not known, but we found a DFT length of $N' = 2^{\lceil \log_2 (N) \rceil}$ to be sufficient to train the models.

\subsection{LFO generator}
The proposed model has an LFO module which governs the time-varying behaviour of the spectral processing. The LFO produces samples at the frame rate, $F_f$, and is defined as the real part of a damped complex exponential:
\begin{equation}\label{eq:lfo_gen}
    s_m(z_a, z_b) = \text{Re}(z_b z_a^m) = |z_b||z_a|^m \cos (m \angle z_a + \angle z_b)
\end{equation}
where $z_a, z_b \in \mathbb{C}$ are the complex frequency and complex amplitude respectively.
Hayes et al. showed that Wirtinger's calculus can be used to compute the partial derivatives of real-valued, complex-variable functions such as \eqref{eq:lfo_gen} to enable sinusoidal frequency estimation by gradient descent \cite{Hayes2022}. In this work we extend this method to include a learnable starting phase and amplitude (defined by $z_b$). The Wirtinger derivatives of \eqref{eq:lfo_gen} are:
\begin{subequations}
\begin{align}
        \frac{\partial s_m}{\partial z_a} \triangleq \frac{1}{2} \left(\frac{\partial s_m}{\partial \text{Re}(z_a)} - j\frac{\partial s_m}{\partial \text{Im}(z_a)}\right) &= \frac{m z_b z_a^{m-1}}{2} \\
    \frac{\partial s_m}{\partial z_b} \triangleq \frac{1}{2} \left(\frac{\partial s_m}{\partial \text{Re}(z_b)} - j\frac{\partial s_m}{\partial \text{Im}(z_b)}\right) &= \frac{z_a^{m}}{2}\,.
\end{align}
\end{subequations}
In the model implementation, the LFO parameters were initialised to:
\begin{equation}
    z_a = 0.7 \exp(j \zeta / F_f), \quad z_b = 1.0
\end{equation}
where $\zeta$ is a random number sampled from a standard normal distribution. The damped amplitude envelope was only applied during training; during inference $z_a$ was normalised to the unit circle to give a lossless LFO. 

\subsection{Multi-layer perceptron waveshaper}
The output of the LFO generator is passed through a multi-layer perceptron (MLP) to allow the model to learn non-sinusoidal control signals like those described in Section \ref{subsec:lfos}. The conceptual motivation for this module was to emulate the linear and non-linear mapping of the LFO signal (in volts or amperes) to the break-frequencies of the all-pass filters (in \unit{\radian\per\second}) in an analog phaser device. Therefore we treat the output of the MLP as the model's prediction of the normalised time-varying break-frequency signal, such that the all-pass parameter is given by:
\begin{equation}
    p_m = \frac{1 - \tan(d_m)}{1 + \tan(d_m)} \quad \text{where} \quad d_m = \text{MLP}(s_m, \gamma)\,,
\end{equation}
with $\gamma$ being the MLP parameters. The MLP consisted of three hidden layers; with 8 neurons per layer; hyperbolic tangent activation functions in the hidden layers; and linear activation in the output layer. The input and output features were scalar, based on the assumption that all $K$ all-pass filters are identical in the reference device. 

\subsection{Model transfer function}\label{subsec:spectral_processing}
The frame-dependent transfer function of the model has the form:
\begin{equation}\label{eq:my_TF}
    \mathbf{h}_m \triangleq \text{diag}(\mathbf{H}_m) = \mathbf{h}^{(1)} \cdot \left(g_1 + \frac{\mathbf{h}^{(2)} \cdot \mathbf{a}_m}{1 - |g_2| \mathbf{z}^{-I(\phi)\phi}\cdot \mathbf{h}^{(2)} \cdot \mathbf{a}_m}\right)
\end{equation}
where  $\mathbf{a}_m$ is the frame-dependent all-pass kernel:
\begin{equation}\label{eq:all_pass_m}
    \mathbf{a}_m = \left(\frac{p_m - \mathbf{z}^{-1}}{1 - p_m \mathbf{z}^{-1}}\right)^K
\end{equation}
and ${\bf z}$ is the $N'$- element vector ${\bf z} = [e^{\frac{2\pi j (0)}{N'}},\hdots,e^{\frac{2\pi j (N'-1)}{N'}}]^{T}$. (Here and elsewhere in this section, operations on vectors are assumed to be applied element-wise.) The parameters $g_1, g_2, \phi \in \mathbb{R}$ retain their physical meanings from Section \ref{sec:phasing} but are now initialised as learnable parameters of the model. $I(\cdot)$ is the Heaviside step function and was included to prevent the model learning a non-casual transfer function. The number of all-pass filters is a hyper-parameter and was fixed for all experiments to $K=4$. The model includes two frequency domain representations of bi-quad filters, given by:
\begin{equation}
    \mathbf{h}^{(i)} = \frac{b_{0}^{(i)} + b_{1}^{(i)} \mathbf{z}^{-1} + b_{2}^{(i)} \mathbf{z}^{-2}}{1 + a_{1}^{(i)} \mathbf{z}^{-1} + a_{2}^{(i)} \mathbf{z}^{-2}}
\end{equation}
where $b_{0}^{(i)}, b_{1}^{(i)}, b_{2}^{(i)}, a_{1}^{(i)}, a_{2}^{(i)} \in \mathbb{R}$ are the learnable filter parameters for the $i$th biquad, $i=1,2$. These kernels were included to account for any further LTI filtering that an analog phaser might impart in addition to the core phasing effect described in Section \ref{sec:phasing}. For example, low-pass filtering, DC blocking or gain adjustment.
\subsection{Loss function}
\label{loss_sec}
The proposed model uses the error-to-signal ratio (ESR) as the objective loss function during training:
\begin{equation}\label{eq:td_loss}
    \mathcal{L}(y, \hat{y}) = \frac{ \sum_{l=0}^{L-1}   \left(y[l] - \hat{y}[l] \right)^2   }
                            {  \sum_{l=0}^{L-1} y[l]^2      }
\end{equation}
where $L$ is the length of the training data in samples. This loss function has been widely used in black-box and grey-box modelling of other audio effects \cite{Wright2020, Wright2022, Wright2021}, but the strict time-alignment required by \eqref{eq:td_loss} introduced some interesting challenges when it came to training the proposed model.
Because the frequency and phase of the LFO are unknown parameters, the training data cannot be arbitrarily split into short segments (as is common when using long audio sequences as training data \cite{Wright2020, Comunita2022}). For example, even if the initial random frequency guess was precisely correct, the model would `see' a different starting phase for each segment (unless the segment length happened to be an integer multiple of the LFO period). In initial experiments, this phase discrepancy was accounted for using the current estimate of LFO frequency, but this caused noisy optimizer updates and convergence issues when the segment length was shorter than one LFO period in the target data.

This presents a dilemma in training this model: we need a sufficient duration of training data to capture the slowly time-varying features in the target system, but are constrained to training with single-batch gradient descent, meaning the time taken for one optimizer step increases linearly with the length of training data. The proposed solution to this problem was to use a short, spectrally-flat training signal. Details of this signal are outlined in Section \ref{sec:target_systems}. 

\subsection{Training details}
All models were trained on audio with a sample rate of \SI{44.1}{\kilo\hertz} using an Adam optimizer \cite{adam2015} with an initial learning rate of $10^{-3}$. Models were trained for a maximum of 5000 epochs on a NVIDIA Titan-X GPU. The training times varied depending on the length of training data and window size. For example, for an audio sequence of \SI{10}{\second} the training times were $\sim$3 hours and $\sim$16 hours for window lengths of \SI{160}{\milli\second} and \SI{10}{\milli\second} respectively.  It is important to note that early designs of the proposed model were implemented in the time-domain and trained via back-propagation through time, but training times were deemed too slow to pursue this approach further ($\sim$24 hours for 1000 epochs on 1s of training audio). The prohibitive training times of IIR filters has been reported in previous work \cite{Nercessian2021} \cite{Wright2022}.

\subsection{Model inference}
In this work, at model inference we use the same algorithm as in training (i.e. using frame-based spectral processing). This has the limitation of introducing a minimum latency of $W$ seconds into the system, which may be unsuitable for real-time use. A time-domain implementation using IIR filters could conceivably be derived via inverse $z$-transform of the system transfer function \eqref{eq:my_TF}, but this is left as a task for future work. The handling of the possibly non-integer delay-line length $\phi$ would require some consideration, but could be implemented with an all-pass filter in the feedback loop shown in Figure \ref{fig:td_model}.

\section{Target systems and datasets}\label{sec:target_systems}
A custom dataset was collected consisting of 60s of a synthetic chirp-train signal followed by 60s of direct-input (DI) guitar recordings. The chirp-train signal was used as the input signal for model training; whereas the guitar recordings were reserved for testing.

\subsection{Synthetic training signal}
The chirp-train signal was synthesised as an impulse train with period \SI{30}{\milli\second} passed through a cascade of 64 all-pass filters \eqref{eq:A(z)} with $p = 0.9$. This type of signal has been used previously for estimating the LFO frequency and shape of LTV audio effects \cite{Wright2021, Kiiski2016}. Due to its spectral flatness, it was hypothesised that even a few seconds of this signal would be sufficient to train the model.

\subsection{Digital phaser}
As a simplified test problem, the proposed model was initially trained on data generated through a digital phaser with transfer function \eqref{eq:tf_z-domain} ($K=4$) implemented through a time-domain recursion in MATLAB. This can be viewed as a specific instance of the model itself but without frame-based spectral processing and under known parameters, shown in Table \ref{tab:dp_params}. The LFO was set to a triangular wave, sweeping through break-frequencies from \SI{4000}{\radian\per\second} to a maximum of \SI{16000}{\radian\per\second}. The maximum \SI{60}{\decibel} decay time of this system (occurring at the minimum of the LFO cycle) was measured to be $t_{60} = $ \SI{38}{\milli\second}.

\subsection{EHX Small Stone}
The Electro-Harmonix (EHX) Small Stone is a commonly encountered analog phaser pedal. The pedal has a single knob to control the rate of the effect, and a binary ``colour'' switch. High-level analysis of a circuit schematic \cite{jdsleep} showed that it consisted of an LFO module, a series of four first order all-pass filter sections and a feedback loop (engaged when the ``colour'' switch is on). This circuit therefore shares the same topology as the proposed model. 
The Small Stone data was collected by processing both the training and testing audio through the pedal in one continuous recording. The audio was sent to the pedal via the output of a PreSonus Audiobox i2 interface, and the output of the pedal re-connected to the input of the audio interface. A calibration recording was obtained with the pedal in bypass-mode and used as the input to the models to negate the effect of the recording equipment on model training. Six unique parameter configurations were captured: three different positions of the `rate' knob with colour switch ON (circuit with feedback) and colour switch OFF (no feedback). The LFO rates were estimated through manual inspection of the spectrogram, providing a pseudo-ground-truth $f_0$ which could later be compared to the learned LFO signals --- see Table \ref{tab:approx_rates}.
\begin{table}[h]
    \centering
    \begin{tabular}{|c|c|c|c|c|}
    \hline
         Label & Colour & Rate knob position & $T_0$* [\unit{\second}] & $f_0$* [\unit{\hertz}] \\
         \hline
         SS-A&  & 3 o'clock & 0.44 & 2.28  \\
         SS-B& OFF & 12 o'clock & 1.60 & 0.625 \\
         SS-C&  & 9 o'clock & 11.6 & 0.086 \\

         \hline
         SS-D &  & 3 o'clock & 0.70 & 1.4 \\
         SS-E & ON & 12 o'clock & 2.56 & 0.38 \\
         SS-F&  & 9 o'clock & 18 & 0.056 \\

          \hline

    \end{tabular}
    \caption{\it{Manually estimated LFO rates of the Small Stone under different parameter configurations. * denotes approximate values.}}
    \label{tab:approx_rates}
\end{table}

\section{Experiments and Results}\label{sec:experiments}
The focus of the experiments presented here is on the effect of the window length $W$ (in seconds) on model accuracy in the context of both model training and inference. In all experiments, the accuracy metric was the resultant ESR \eqref{eq:td_loss} on the test dataset. It was noted that the training convergence was sensitive to the initialisation of the MLP parameters, $\gamma$. To address this issue, each training procedure was re-initialised and repeated three times. The iteration with the lowest ESR was retained. A frame overlap of 75\% was used across all experiments, with the frame length in samples $N$ truncated to a multiple of four to ensure constant-overlap-add \cite{PASPWEB2010}. 

\subsection{Experiment 1: training frame size sweep}
As an initial experiment, instances of the model were trained with frame lengths ranging from \SI{10}{\milli\second} to \SI{160}{\milli\second} on the following data:
\begin{enumerate}[label=(\alph*)]
    \item Digital phaser with LFO rate $T_0=$ \SI{2}{\second} (DP-2).
    \item Small Stone with parameter configuration A (SS-A). 
    \item Small Stone with parameter configuration D (SS-D).
\end{enumerate}
The training data was truncated to \SI{2.67}{\second} in duration to accelerate training, and was deemed sufficient given that it contained at least one LFO cycle for all case studies.

\begin{figure}[h]
    \centering
    \includegraphics[scale=0.65, clip, trim=0mm 21mm 5mm 0mm]{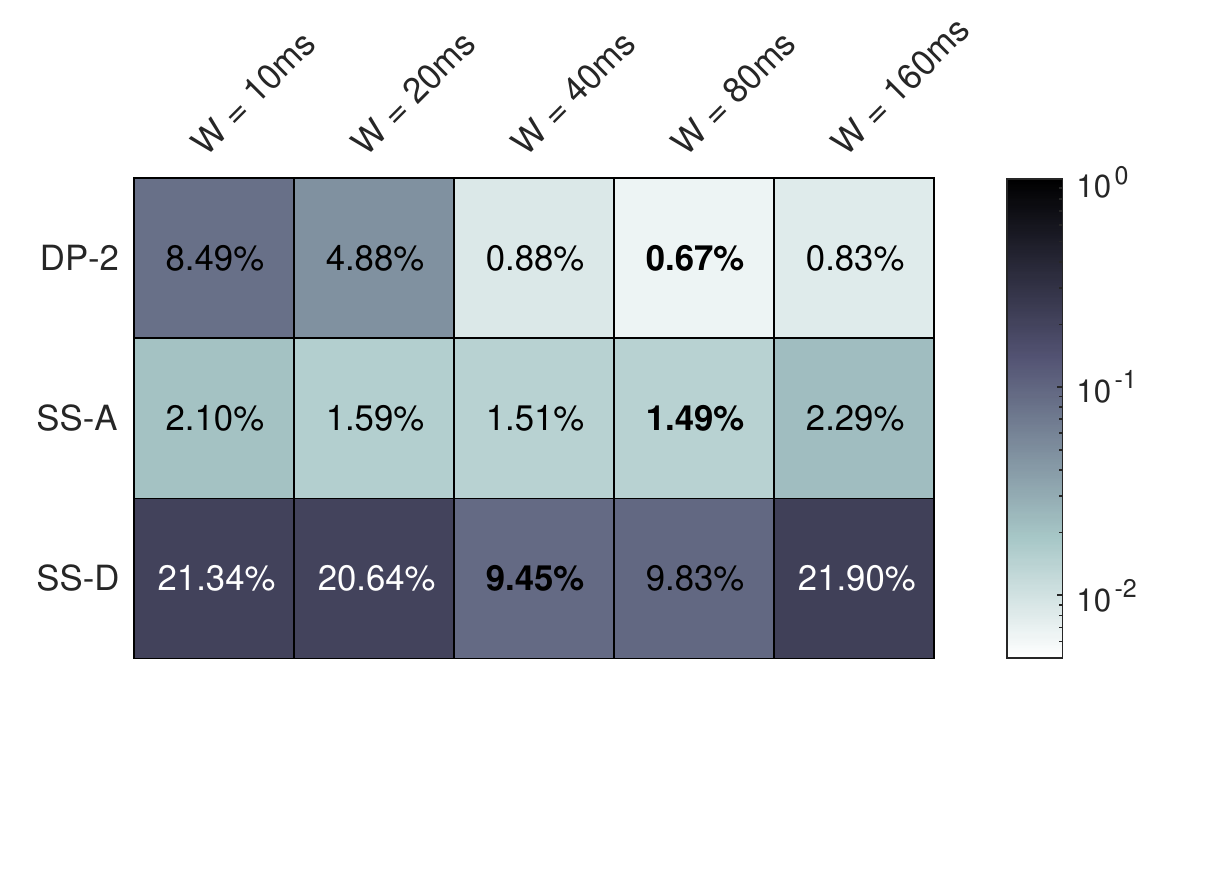}
    \caption{\it{ESR for different training window lengths $W$ on the test audio of three case studies: digital phaser with $T_0=2$s (DP-2), Small Stone with parameter configurations A and D (SS-A, SS-D).}}
    \label{fig:exp1-color-table}
\end{figure}

\begin{figure*}
        \centerline{\includegraphics[scale=0.5, clip, trim=0mm 0mm 0mm 0mm]{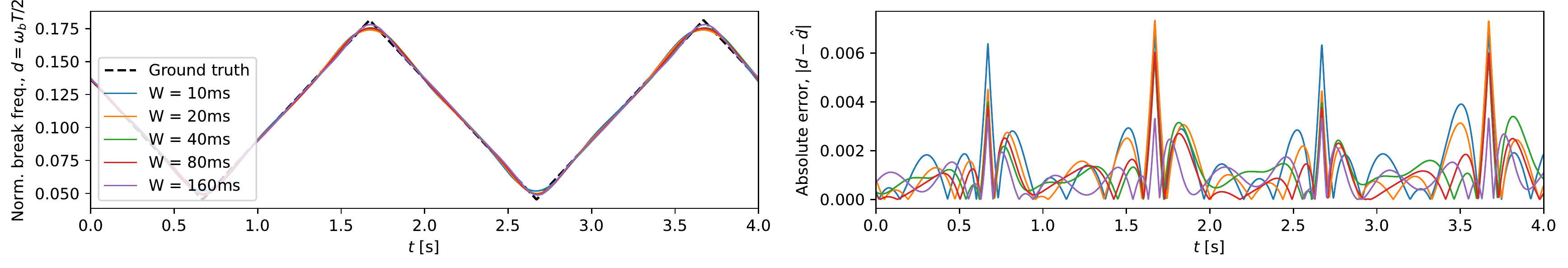}}
        \caption{\label{fig:waveshapes_dp}{\it Outputs of the MLP module (left) and the absolute error (right) compared to the triangular LFO in the target digital phaser, with $T_0 = 2$s (plotted left as ground truth).}}
\end{figure*}

The ESR obtained in experiment 1 can be seen in Figure \ref{fig:exp1-color-table}. In the case of the digital phaser, frame lengths of \SI{40}{\milli\second} to \SI{160}{\milli\second} all resulted in a error-to-signal of less than 1\% on the testing data -- implying an accurate match between the model output and target waveform with minimal artefacts introduced by windowing. The frame lengths of \SI{10}{\milli\second} and \SI{20}{\milli\second} produced worse results, 
suggesting an insufficient number of bins in the transfer function shape the spectrum and/or severe time aliasing in the output. The learned parameters during this experiment can be found in Table \ref{tab:dp_params} and show a good estimation of the parameters in the target model. Figure \ref{fig:waveshapes_dp} shows the synthesised LFO signals, compared to the target triangular wave.

\begin{figure*}
        \centerline{\includegraphics[scale=0.5, clip, trim=0mm 0mm 0mm 0mm]{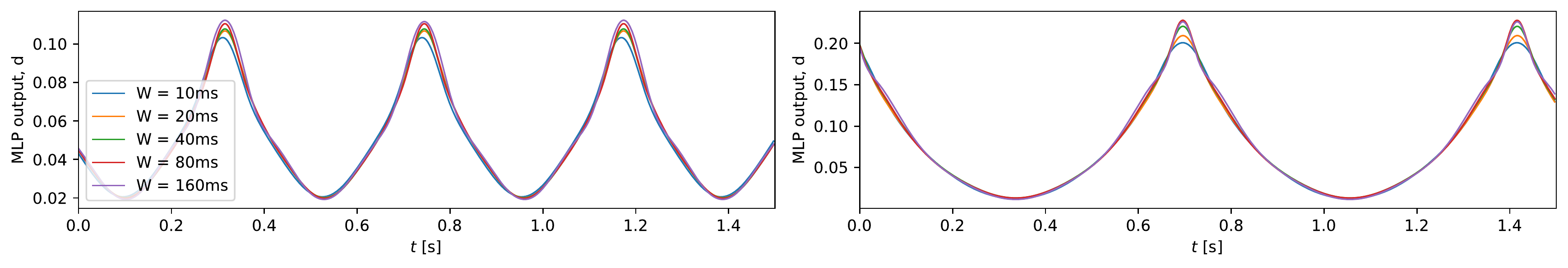}}
        \caption{\label{fig:waveshapes_ss}{\it Outputs of the MLP module in the Small Stone modelling task for different training window lengths with feedback off (SS-A, left) and feedback on (SS-D, right). In both cases, the training audio was recorded with the pedal's rate knob at 3 o'clock. NB the ground-truth signal is unknown so not plotted.}}
\end{figure*}

In the case of the Small Stone, the resulting model accuracy depended on the presence of feedback in the circuit. The minimum test loss was approximately 1.5\% without feedback and 10\% with feedback. This result is expected due to the increase in circuit complexity and longer decay time associated with the feedback case. In both cases, window lengths of \SI{40}{\milli\second} and \SI{80}{\milli\second} provided the best performance---suggesting a good trade-off between time and frequency resolution. Despite the discrepancies in numerical results, the perceptual differences are difficult to distinguish, informally, from the target system---however, some differences in the low-frequencies are noted for short window lengths. The reader is referred to the accompanying web-page for audio examples. It is interesting to observe the learned LFO signals of the Small Stone, as shown in Figure \ref{fig:waveshapes_ss}. In both cases, the MLP module has consistently predicted a similar wave-shape across the frame-rates. When engaged, the colour switch appears to increase both the depth and the period of break-frequency modulation.

    \hfill

\subsection{Experiment 2: training frame size vs LFO rate}
Experiment 2 investigated the effect of frame-length on model accuracy in more detail, considering as a case study the digital phaser with LFO periods $T_0=$ \qtylist{0.5; 2; 8}{\second} The length of training data was held constant at \SI{10}{\second}, and the prior knowledge of target LFO periods informed the choice of frame-lengths:
\begin{equation}
    W_b = T_0 2^{b/2} / 100 \quad {\rm where}\quad b = 0, \dots, 10
\end{equation}

\begin{figure}[h]
\centerline{\includegraphics[scale=0.6, clip, trim=0mm 0mm 0mm 0mm]{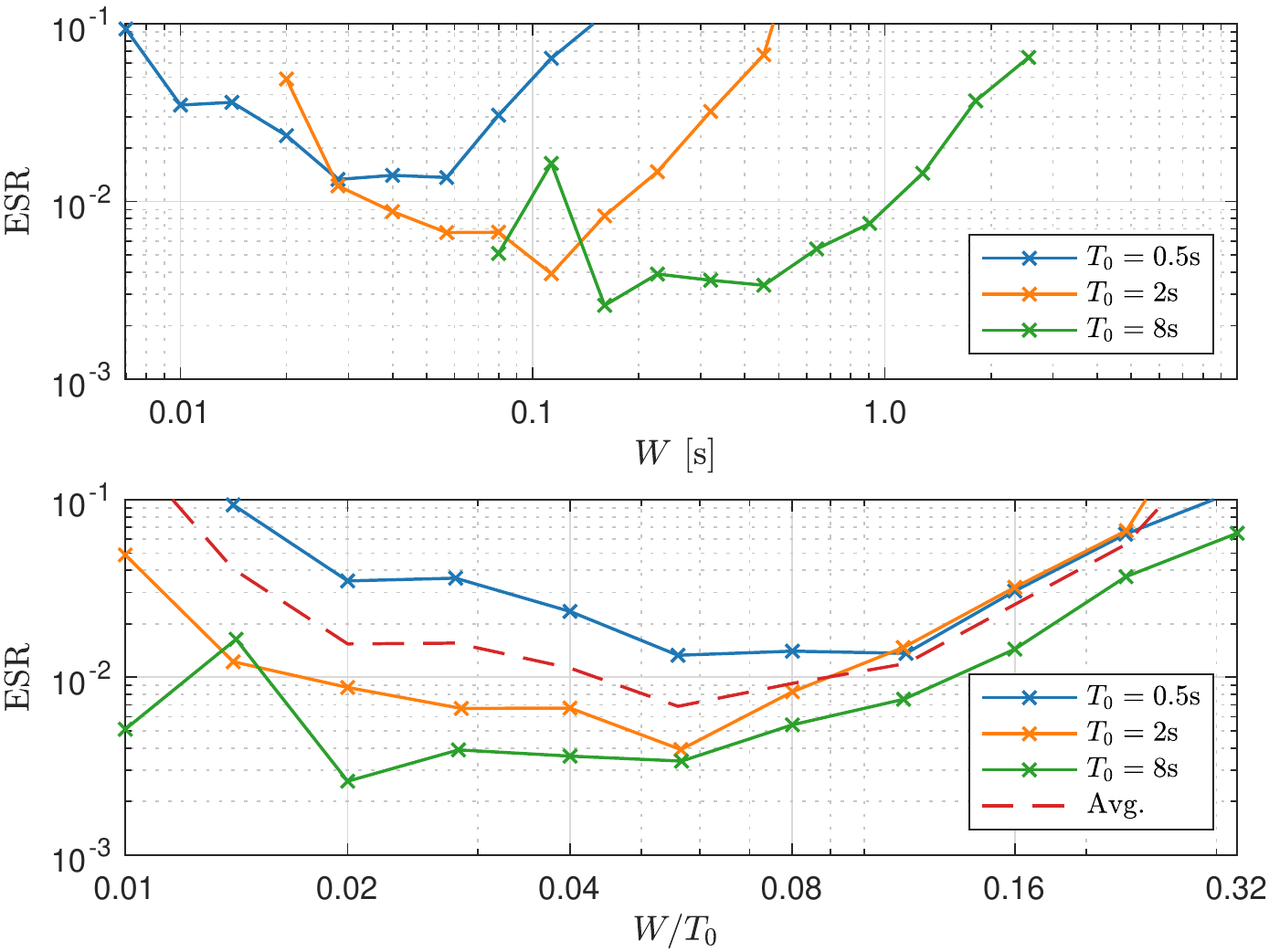}}
\caption{\label{fig:exp2-results}{\it ESR against training window length $W$ (top) and the ratio of training window length to target LFO period $W/T_0$ (bottom) for the digital phaser.}}
\end{figure}

Figure \ref{fig:exp2-results} (top) shows the results of the experiment, with minimum test loss against training frame size for different rates of phaser effect. Firstly, we see that the model accuracy increases for longer LFO periods. This is intuitive, as in the limit $T_0 \to \infty$ the target system becomes linear and time-invariant (LTI) so we expect the artefacts of frame-based processing to diminish. Also intuitively, the results suggest the optimum window length depends on the target LFO period. In the bottom figure, the window length has been normalised to target LFO period. In this case, the optimum $W/T_0$ ratio shows consistency across the phaser rates, with $W/T_0 \approx 5\%$ giving, on average, the lowest loss.

\subsection{Experiment 3: inference frame size}
\begin{figure}[h]
\centerline{\includegraphics[scale=0.65, clip, trim=5mm 33mm 10mm 10mm]{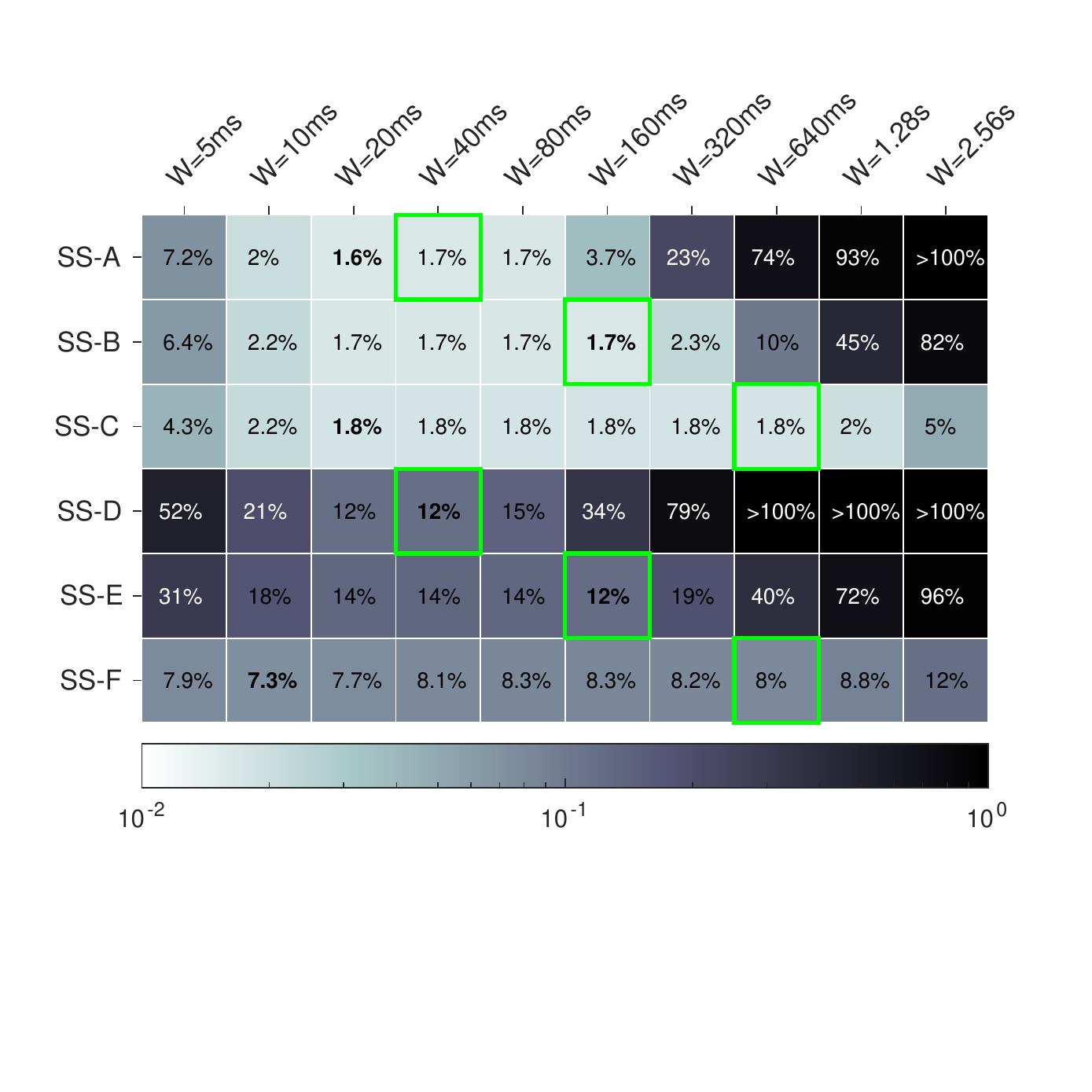}}
\caption{\label{fig:exp3-results}{\it ESR using various window sizes at inference time across the Small Stone parameter configurations. The green squares indicate the training window lengths.}}
\end{figure}

The aim of the final experiment was two-fold: to train instances of the model on all six parameter configurations of the Small Stone, and to investigate the effect of window length on model accuracy during inference. For each configuration, the training window length was informed by the results of Experiment 2 and set within 5-10\% of the estimated LFO period (see Table \ref{tab:approx_rates}). The training data was truncated to contain approximately three cycles of the LFO. After training, the models were tested using various window lengths at inference, with the results shown in Figure \ref{fig:exp3-results}

In both feedback configurations, the models trained on higher LFO rates (SS-A, SS-D) were most sensitive to changes in window size at inference time. This implies a fine balance between the window size being long enough to simulate the transient behaviour of the device, but short enough to not smear the LFO behaviour. In contrast, the accuracy of models trained on longer LFO periods (SS-C, SS-F) was mostly unchanged across the range of window sizes tested. This is a promising result, as it implies one can use a long window size for accelerated training; but a short window size for lower-latency playback at inference. However, the results suggest that there will always be a lower bound on the frame size that is determined by the decay time of the system modelled. 

\begin{table*}[h]
    \centering
    \begin{tabular}{|l|l|l|l|l|}
    \hline
        Parameter &  Description & Target value & Initial Value & Learned Value (to 3.s.f)\\
        \hline\hline
         $f_0$ & LFO rate [Hz] & 0.5 & 0.0627 & 0.500 \\\hline
         $g_1$ & Wet mix & 1.0 & 1.0 & 0.999 \\\hline
         $g_2$ & Feedback gain & 0.7 & 0.01 & 0.700 \\\hline
         $\phi$ & Feedback delay-line length (samples) & 1 & 0.5 & 0.995 \\\hline
         $[b_{0_1}, b_{1_1}, b_{2_1}]$ & Biquad 1 feedforward coeffs. & [1, 0, 0] & [1, 0, 0] & [1.00,   -0.0641,   0.0336]\\\hline
        $[a_{1_1}, a_{2_1}]$ & Biquad 1 feedback coeffs. & [0, 0] & [0, 0] & [-0.0629, 0.0336] \\\hline
         $[b_{0_2}, b_{1_2}, b_{2_2}]$ & Biquad 2 feedforward coeffs. &  [1, 0, 0] & [1, 0, 0] & [1.00,   -0.0214,    0.0140] \\\hline
        $[a_{1_2}, a_{2_2}]$ & Biquad 2 feedback coeffs. & [0, 0] & [0, 0] & [-0.0238, 0.0136] \\
        \hline
    \end{tabular}
    \caption{Example learned parameters of digital phaser model with $T_0 = $ \SI{2}{\second} (DP-2). The parameters are from the best performing model in Experiment 1, obtained using a window size of 80ms during training. Note that the biquad feedforward coefficients have been normalised.}
    \label{tab:dp_params}
\end{table*}

\section{Conclusions and further work}\label{sec:conclusion}
This work has presented a differentiable DSP model of a phaser that uses frame-based spectral processing to implement a time-varying filter in the frequency domain. The model was based on a generalised continuous-time model of a phaser effect with several free parameters learnable via gradient descent. It was shown that the model can recover the parameters of a reference digital phaser and learn the correct frequency, starting phase and waveform of the underlying low-frequency oscillator (LFO), without seeing the ground-truth LFO during training. Furthermore, the model was trained to emulate an analog reference device. Informal listening found the model perceptually convincing in this task for a range of parameter configurations. Formal listening tests are important, but left for future work. It was found that the objective model accuracy depended on the training window length and required a manual estimation of the target LFO frequency for the best results. Future work will aim to remove the need for this initial estimation, perhaps through a multi-resolution training process. A key limitation of the proposed model is the inherent latency introduced by the frame-based approach, which could be problematic for real-time model inference. Future work will aim to remove this by implementing an equivalent audio-rate time-domain recursion. Finally, further work may involve extending the general approach proposed in this paper to grey-box modelling of other time-varying filters and delay-based audio effects such as auto-wah, flangers and chorus.

\section{Acknowledgments}
The authors would like to thank Ben Hayes, Lauri Juvela and Alec Wright for helpful discussions on some underlying concepts of this work. Thank you to the anonymous reviewers for their comments.

\begin{scriptsize}
\bibliographystyle{IEEEbib}
\bibliography{DAFx23_tmpl} 
\end{scriptsize}
\end{document}